\documentclass[aps,prd,a4paper,twocolumn,showpacs,showkeys,preprintnumbers,amsmath,amssymb]{revtex4}
\usepackage{graphicx}
\usepackage{bm}
\usepackage{dcolumn}
\usepackage{color}
\usepackage{pst-plot}
\renewcommand{\texttt}{{}}
\newcommand{\be}{\begin{eqnarray}}
\newcommand{\ee}{\end{eqnarray}}
\usepackage{graphicx}

\begin{document}

%\title{The microstructure of a quantum spacetime}
\title{Spectral dimension of a quantum universe} % of Loop Quantum Gravity}
\author{Leonardo Modesto}
\thanks{Electronic address: lmodesto@perimeterinstitute.ca}
\affiliation{Perimeter Institute for Theoretical Physics, 31 Caroline Street, Waterloo, ON N2L 2Y5, Canada}

\author{Piero Nicolini}
\thanks{Electronic address: nicolini@th.physik.uni-frankfurt.de}
\affiliation{Frankfurt Institute for Advanced Studies (FIAS),
%Johann Wolfgang Goethe University
Institut f\"ur Theoretische Physik, Johann Wolfgang Goethe-Universit\"at, Ruth-Moufang-Strasse 1,
60438 Frankfurt am Main, Germany}

\date{\small\today}

\begin{abstract} \noindent
In this paper, we calculate in a transparent way the spectral dimension of a quantum spacetime, considering a diffusion process propagating on a fluctuating manifold. To describe the erratic path of the diffusion, we implement a minimal length by averaging the graininess of the quantum manifold in the flat space case. As a result we obtain that, for large diffusion times, the quantum spacetime behaves like a smooth differential manifold of discrete dimension. On the other hand, for smaller diffusion times, the spacetime looks like a fractal surface with a reduced effective dimension. For the specific case in which the diffusion time has the size of the minimal length, the spacetime turns out to have a spectral dimension equal to 2, suggesting a possible renormalizable character of gravity in this regime. For smaller diffusion times, the spectral dimension approaches zero, making any physical interpretation less reliable in this extreme regime. We extend our result to the presence of a background field and curvature. We show that in this case the spectral dimension has a more complicated relation with the diffusion time, and conclusions about the renormalizable character of gravity become less straightforward with respect to what we found with the flat space analysis.
\end{abstract}

\maketitle
\begin{figure}
 \begin{center}
 \includegraphics[height=8.5cm]{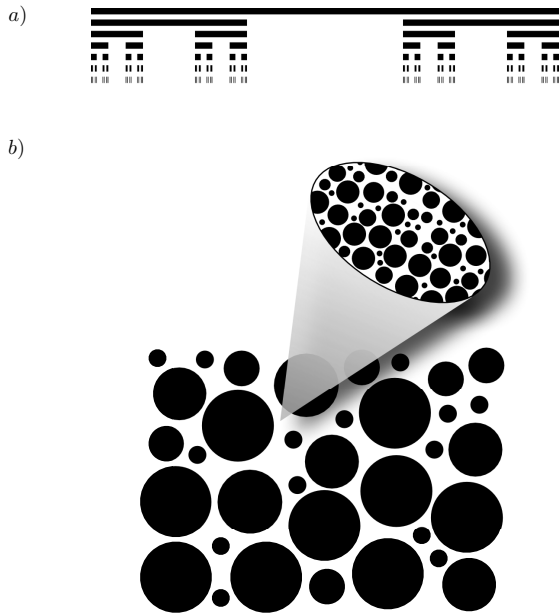}
 \hspace{0.2cm}
      \caption{\label{Plot0}In (a) we have a Cantor set, i.e. a fractal surface which resembles the quantum spacetime. Probing the spacetime with increasing energy corresponds to moving from the upper continuous surface to the lower fractal ones. In (b), there is an artistic picture of the spacetime, whose holed structure shows a fractal self-similarity.
 }
 \end{center}
  \end{figure}

One of the most astonishing features of quantum gravity is the supposed emergence of a new dynamical variable, which replaces the conventional concept of spacetime dimension. A possible explanation for this conjecture would lie in the fact that the spacetime exhibits graininess at energy scales at which its classical description in terms of smooth differential manifold breaks down. In other words, we are dealing with something like a surface that at large distances (low energy) appears smooth, while at close distances (high energy) it is rough. For this reason, in the quantum gravity regime one not only loses any resolution of pointlike structures,  but also has a modified perception of any topological dimension, which classically we take for granted.
For instance, in the context of quantum string loops, the Hausdorff dimension of the string world sheet is an actual indicator of the amount of fuzziness perceived by a detecting apparatus. In this scenario, the spacetime, in its classical picture, emerges as a $p$-brane condensate, which can evaporate, i.e. make a transition to an excited configuration as far as length scales of order $(\alpha^\prime)^{1/2}$ are concerned. In this case the world sheet becomes a fractal surface of dimension three, since the energy in the excited state lets the string explore an additional dimension \cite{Ansoldi:1997cw,Ansoldi:1998ys,Aurilia:2002aw}.
Conversely, in the context of a diffusion process or a random walker on the spacetime manifold, one discovers that the blurriness of the spacetime at short distances screens some dimensions, reducing the number of those actually accessible. In this case the detecting apparatus perceives again the spacetime in terms of a fractal surface, whose spectral dimension ${\mathbb D}$ smoothly  approaches the discrete topological  value $d$ as long as the diffusion/walking time increases and  larger scales are probed \cite{Carlip:2009kf} (see Fig. (\ref{Plot0}). The virtue of such effective dimensional reduction lies in the possibility for gravity to be a renormalizable theory, manifesting an ``apparent'' nonrenormalizable character only in its classical (large distance) $d$-dimensional picture.
As a key point of the above considerations there is the mechanism which provides the fuzziness of spacetime in its quantum regime. To this purpose, calculations have been performed in the context of loop quantum gravity (LQG)
\cite{Ambjorn:2005db,Modesto:2008jz,Modesto:2009kq,Caravelli:2009gk,Magliaro:2009if}, asymptotically safe gravity (ASG)
\cite{Lauscher:2005qz},  and $\kappa$-Minkowski noncommutative spacetime ($\kappa$-NC) \cite{Dario}. Even if the spectral dimension is not the unique generalization of dimension, the above calculations provide promising  signals, since at short distances the spectral dimension might converge to two, ${\mathbb D}=2$, where gravity is known to be renormalizable. For instance, both in LQG and ASG the spectral dimension runs from two to four, when the energy scale goes from Planckian to lower ``classical'' values. For $\kappa$-NC, the spectral dimension has been calculated both for the toy model of a quantum sphere and the $\kappa$-Minkowski space with $\kappa=1$, but never reaches the value ${\mathbb D}=2$.  However, in all the above cases the results have been obtained with some numerical integration and/or approximations, while it would be of vital importance to get a more transparent relation between the spectral dimension and the diffusion time. To address this problem we will follow another route employing a quantum spacetime in which an effective minimal length emerges as the average of quantum geometry fluctuations \cite{Cho:1999sg,Smailagic:2003yb,Smailagic:2003rp,Smailagic:2004yy,Spallucci:2006zj,Banerjee:2009gr,Banerjee:2009xx}.
In this picture, the blurriness of the quantum spacetime has a quasiclassical countenance, inducing a smearing effect of any pointlike object.  This approach has already been successfully employed in a variety of contexts, namely to improve the singularity of conventional black hole solutions \cite{Nicolini:2005de,Nicolini:2005zi,Nicolini:2005vd,Rizzo:2006zb,Ansoldi:2006vg,Spallucci:2008ez,Arraut:2009an,Nicolini:2009gw}, to get a thermodynamically stable final stage of the Hawking evaporation \cite{DiGrezia:2006rw,Di Grezia:2007uy,Casadio:2008qy} (for a review see Ref. \cite{Nicolini:2008aj} and the references therein), to describe a traversable wormhole sustained by quantum geometry fluctuations \cite{Garattini:2008xz}, to remove the initial cosmological singularity and drive the inflation without an inflaton field \cite{Rinaldi:2009ba}, and to get corrections to the Unruh thermal bath by means of a nonlocal deformation of conventional field theories \cite{Casadio,Nicolaevici,Nicolini:2009dr}.

We now briefly recall the concept of spectral dimension as it emerges in a diffusion process. We consider a $d$-dimensional Euclidean geometry and the associated heat equation
\begin{eqnarray}
\Delta K\left( x, y ; s \right)=
\frac{\partial}{\partial s}\,
K\left( x, y ; s \right)
\end{eqnarray}
where $s$ is a fictitious diffusion time of dimension of a length squared, $\Delta$ is the Laplace operator, and $K\left( x ,y ; s \right)$ is the heat kernel, representing the probability density of diffusion from $x$ to $y$. As initial conditions, we have that the probability is peaked at the starting point $x$,
\be
K\left( x ,y ; 0 \right)=\frac{\delta^d (x-y)}{\sqrt{\det g_{ab}}}
\ee
where $\delta^d (x-y)$ is the $d$-dimensional Dirac delta and $g_{ab}$ is the metric of the manifold. It is worthwhile to introduce the so-called return probability
\be
P_g(s)=\frac{\int d^d x\sqrt{{\mathrm \det} \ g_{ab}}\ K(x,x;s)}{\int d^d x\sqrt{{\mathrm \det}\ g_{ab}}}
\ee
and define the spectral dimension as
\be
{\mathbb D}=-2 \, \frac{\partial \ln P_g(s)}{\partial \ln s}.
\label{spectrad}
\ee
It is easy to show that in flat space, for a ``free'' diffusion, the return probability is $P_g(s)=(4\pi s)^{-d/2}$ and the spectral dimension is ${\mathbb D}=d$. In the presence of background fields or gravity, the above formula can still be employed to check an effective dimensional reduction, even if the large $s$ limit holds on (finite) patches of the manifold only.

We are now ready to implement an effective minimal length in our manifold. As a start we will address the free diffusion in a blurring background in the absence of gravity. In other words, we would like to understand what the response of the spacetime fabric is, once it is probed by the high energy regime of the diffusion process. Since this regime takes place for the initial values of the diffusion time, we expect that the primary disturbance emerging from the quantum manifold affects the profile of the initial configuration. As a result, advancing in diffusion time, the additional disturbances turn out to be of minor importance. Given this scenario, the heat equation
\begin{eqnarray}
&&\Delta\, K_\ell\left( x ,y ; s \right)=
\frac{\partial}{\partial s}\,
K_\ell\left( x ,y ; s \right)\nonumber\\
&& K_\ell\left( x ,y ; 0 \right)=\rho_\ell\left( x ,y \right) \label{heat}
\end{eqnarray}
coming from averaging the spacetime fluctuations \cite{Spallucci:2006zj} fully meets the physical conditions of the system. Here the function
\be
\rho_\ell\left( x ,y \right)=\left(\frac{1}{ 4\pi \ell^2 }\right)^{\frac{d}{2}} \ e^{-\left( x-y \right)^2/4\ell^2}
\ee
is nothing but a Gaussian distribution, whose width coincides with the minimal length $\ell$ implemented in the manifold.  In other words, the Gaussian, the most narrow allowable distribution, is sustained by the spacetime fluctuations which prevent its collapse into a pointlike Dirac delta.
The resulting heat kernel is
\be
K_\ell\left( x ,y ; s \right)=\frac{e^{- \frac{\left( x-y \right)^2}{4 (s + \ell^2)
 } }}{\left[  4\pi \left(s+\ell^2 \right) \right]^{d/2} }
\ee
which still has a Gaussian profile for all $s$, as expected, because the more $s$ increases, the less fluctuations are important. Now, employing Eq.(\ref{spectrad}), we find that the spectral dimension is
\begin{eqnarray}
{\mathbb D}= \frac{s}{s+\ell^2}\ d.
\label{piatto}
\end{eqnarray}
To start commenting on this result, we see that for large $s$ the spectral dimension ${\mathbb D}$ has the wished  discrete topological value $d$. This can be interpreted by saying that at large distances with respect to $\ell$, the quantum disturbances  are rather weak and the spacetime ultimately behaves like a smooth differential manifold. On the other hand, for any finite value of $s>\ell^2$, the spacetime starts hiding some of its dimensions due to its blurriness, and we get ${\mathbb D}<d$. For smaller $s$, the scenario becomes even more interesting: the spacetime fabric powerfully enters the scene as $s$ approaches $\ell^2$. For the specific case in which the length scale probed by the diffusion process coincides with the minimal length, i.e. $s=\ell^2$, we are in the full quantum regime and we discover that quantum gravity is a renormalizable theory since the spectral dimension is ${\mathbb D}=2$. For smaller $s$, i.e. $s<\ell^2$, we have that the spectral dimension further decreases, ${\mathbb D}<2$, vanishing for $s=0$.  Having a spacetime with less than two dimensions might be difficult to interpret, but at trans-Planckian energies even the very notion of spacetime, as conventionally known, becomes ill defined. We argue that in the regime $s<\ell^2$ the spacetime completely dissolves, assuming a new nature whose analysis goes far beyond the goal of this paper.

The above scenario can be drawn from another, more illuminating perspective. Inverting Eq. (\ref{piatto}), we obtain a new ``clock'' for our diffusion process,
\begin{eqnarray}
s= \frac{{\mathbb D}}{d-{\mathbb D}}\ \ell^2,
\end{eqnarray}
since the spectral dimension can now be used both to drive the diffusion and to control the amount of fuzziness in terms of the ``dimensional evolution'' of the spacetime.
The heat equation now reads
\begin{eqnarray}
&&\Delta K_\ell\left(x, y; {\mathbb D}\right)=\frac{(d-{\mathbb D})^2}{d \, \ell^2}
\frac{\partial}{\partial {\mathbb D}}\,
K_\ell \left(x, y;  {\mathbb D} \right)
\end{eqnarray}
with initial conditions as in Eq. (\ref{heat}).
The solution can now be cast in the form
\be
K_\ell \left( x ,y ; {\mathbb D} \right)=\left(\frac{d - {\mathbb D}}{ 4\pi  d \, \ell^2 }\right)^{\frac{d}{2}} \ e^{-\left( x-y \right)^2(d-{\mathbb D})/4d  \ell^2}.
\ee
The diffusion starts in a trans-Planckian regime and the blurriness is so strong that the thing that plays the role of what we call spacetime has a spectral dimension ${\mathbb D}=0$. As ${\mathbb D}$ increases to 2, the fuzziness decreases, the Gaussian profile of the heat kernel becomes wider, and gravity is a super-renormalizable theory. Only at the Planck scale, for ${\mathbb D}=2$, does gravity become a renormalizable theory. From now on, i.e. ${\mathbb D}>2$, the fuzziness further decreases, the spacetime approaches the status of a $d$-dimensional differential manifold, the probability density becomes constant, and gravity is no longer renormalizable.

For completeness, it is worthwhile to compute the Green's function as
\be
G_\ell \left( x ,y  \right)=\ell^2d\int_0^d \ \frac{d{\mathbb D}}{(d-{\mathbb D})^2}\, K_\ell\left( x ,y ; {\mathbb D} \right)
\ee
which, once integrated, reads
\be
G_\ell \left( x ,y \right)=\frac{\gamma\left((d-2)/2;  \Delta x^2/4\ell^2\right)}{(4\pi)^{d/2} (\Delta x^2)^{(d-2)/2}}
\ee
where $\Delta x^2=(x-y)^2$ and
\be
\gamma\left( a/b ;  x  \right)=\int_0^x \frac{du}{u} u^{a/b} e^{-u}
\ee
is the lower incomplete Euler function.
%% The coincidence limits, $\Delta x\to 0$, are for $d>2$
%% \be
%% G_\ell\left( x ,y \right)\simeq-\frac{1}{2^{d-1}\pi^{d/2}(d-2)\ell^{d-2}}
%% \ee
%% while for $d=2$
%% \be
%% G_\ell\left( x ,y \right)\simeq-\frac{1}{4\pi}\ln(1/\mu^2\ell^2)
%% \ee
%% and finally for $d<2$
%% \be
%% G_\ell\left( x ,y \right)\simeq\frac{2}{2-d}\left[ \ell^{2-d}-\mu^{d-2}\right],
%% \ee
%% where $\mu$ is a massive IR regulator.
In the coincidence limit, the Green's function shows UV regularity for any dimension $d$.
%%  while the appearance of some malicious IR behavior has nothing to do with the procedure we followed, but it is only connected to the reduction of the spacetime dimension and naturally arises in (super)-renormalizable regimes.

The diffusion process can also be studied in the presence of a background field or gravity. In the former case, we have
\begin{equation}
K_\ell\left( x , x ; s \right)= \frac{a_0 + e^{ \left[ \ell^2  s/\left(
\ell^2 +s \right)\,\right]\, \Delta}\, \sum_{n=1}^\infty  s^n
a_n \left( x \right)}{\left[
(4\pi)\,\left( s+\ell^2 \right) \right]^{d /2} }
\label{trace}
\end{equation}
where the nonlocal operator $e^{\left[ \ell^2 s/\left(\ell^2 +s\,\right) \right] \Delta}$ determines the smearing of the coefficients $a_n\left( x \right)$ \cite{Spallucci:2006zj}, which account for the coupling with a generic Abelian background gauge field $A(x)$.  To perform the calculation of the spectral dimension ${\mathbb D}_{f}$, we have to follow the procedure suggested in
\cite{Ambjorn:2005db,Lauscher:2005qz}, invoking a parameter asymptotic ordering. We assume that the diffusion parameter $\sqrt{s}$ can still be larger with respect to $\ell$, but always small with respect to $L$, the characteristic dimension of the manifold or of one of its patches where the diffusion takes place.
As a result, the spectral dimension can again be calculated without approximations. Here, for the sake of clarity, we provide the primary corrections only just to say that Eq. (\ref{piatto}) holds up to subleading terms, namely,
\begin{eqnarray}
&&{\mathbb D}_{f}\simeq {\mathbb D} + \frac{s}{a_0}\left(\frac{sd}{s+\ell^2}-2\right) \int e^{\left[ \ell^2 s/ \left(
\ell^2 +s \right)\right] \Delta}  a_1\left( x \right)+\nonumber\\&&-\frac{2s^2}{a_0} \left(\frac{\ell^2}{\ell^2+s}\right)^2 \int e^{\left[ \ell^2 s/\left(
\ell^2 +s \right) \right] \Delta} \Delta   a_1 \left( x \right)+\dots
\label{campo}
\end{eqnarray}
provided that $a_n$ are regular. The integral symbol $\int$ stands for the integral operator acting on a generic integrable function $f(x)$ as $[\int d^d x\sqrt{{\mathrm \det} \ g_{ab}}\ f(x)]/\int d^d x\sqrt{{\mathrm \det}\ g_{ab}}$, while ${\mathbb D}$ is the result found in (\ref{piatto}).

In the presence of gravity, we have to consider further contributions due to the presence of curvature effects.
As a start, the Laplace operator acquires an extra term accounting for the Ricci scalar,
\begin{equation}
\Delta\longrightarrow \Delta_g\equiv \Delta -\xi_d \, R \ , \qquad \xi_d\equiv
\frac{1}{4}\frac{d-2}{d-1}
\end{equation}
while the minimal length, being an invariant scalar, gets, in general, a spacetime dependence,
$\ell\longrightarrow \ell\left(\, x\,\right)$. As a result the heat kernel reads \cite{Spallucci:2006zj}
\begin{eqnarray}
&&K_\ell\left( x , x ; s \right)=\\
&&=\frac{  a_0 e^{ \frac{s\ell^2\left( x \right)}{ s+\ell^2
\left( x \right) }  \xi_d R} + e^{ \frac{s \ell^2 \left( x \right)}{ s+\ell^2\left(x
\right) }
    \Delta_g} \sum_{n=1}^\infty\, s^n   a_n\left(
x ,x \right) }{\left[ \left(\, 4\pi\,\right)
\left( s+\ell^2\left( x \right) \right) \right]^{d/2} }\,\nonumber
\end{eqnarray}
and the spectral dimension gets additional terms, coming even from the $a_0$ coefficient. Again we provide the leading corrections only, i.e.,
\begin{eqnarray}
&&{\mathbb D}_g\simeq \int_R \ {\mathbb D}(x)-2s\int_R\left(\frac{\ell^2(x)}{\ell^2(x)+s}\right)^2\xi_d R+\label{curvo}\\
&&\frac{2s}{a_0}\int_{\Delta_g}\left[a_1(x)+\left(\frac{\ell^2(x)}{\ell^2(x)+s}\right)^2\Delta_g a_1(x)-\frac{s\ d/2}{s+\ell^2(x)}a_1\right]\nonumber
\end{eqnarray}
where ${\mathbb D}(x)\equiv  s d/(s+\ell^2(x))$, while for a generic integrable function $f(x)$,
\be
\int_R f(x)\equiv\frac{\int d^d x\frac{\sqrt{{\mathrm \det}\ g_{ab}}}{(s+\ell^2(x))^{d/2}}\ e^{ \frac{s\ell^2\left( x \right)}{ s+\ell^2
\left( x \right) }  \xi_d R}\ f(x)}{\int d^d x\frac{\sqrt{{\mathrm \det}\ g_{ab}}}{(s+\ell^2 (x))^{d/2}}\ e^{ \frac{s\ell^2\left( x \right)}{ s+\ell^2
\left( x \right) }  \xi_d R}}
\ee
and
\be
\int_{\Delta_g} f(x)\equiv\frac{\int d^d x\frac{\sqrt{{\mathrm \det}\ g_{ab}}}{(s+\ell^2(x))^{d/2}}\ e^{ \frac{s\ell^2\left( x \right)}{ s+\ell^2
\left( x \right) }   \Delta_g}\ f(x)}{\int d^d x\frac{\sqrt{{\mathrm \det}\ g_{ab}}}{(s+\ell^2 (x))^{d/2}}\ e^{ \frac{s\ell^2\left( x \right)}{ s+\ell^2
\left( x \right) }  \xi_d R}}.
\ee
We notice that in both Eqs. (\ref{campo}) and Eq. (\ref{curvo}), the spectral dimension vanishes for $s\to 0$. For large $s$ one recovers the discrete value $d$, using the asymptotic ordering $s/L^2\ll 1$ as in \cite{Ambjorn:2005db,Lauscher:2005qz}. Indeed, even in the presence of gravity one can prove that only the first term on the right-hand side of Eq. (\ref{curvo}) survives. Since invoking the asymptotic ordering is equivalent to having a ``local'' (flat space) diffusion, we can conclude that, actually, $\int_R {\mathbb D}(x)\to d$. The case $s=\ell^2$ is again very interesting, because
 the presence of a background field and/or curvature further disrupts the spacetime, and the value of the spectral dimension strongly fluctuates around $2$. This implies that the possibility for gravity to be  a renormalizable theory within an effective dimensional reduction is subject to nontrivial corrections coming from all terms in the Seeley-deWitt series for the heat kernel \cite{Ahmed:2009qm,Benedetti:2009ge}.

In this paper, we presented a new approach to compute in a clear way the spectral dimension of a quantum spacetime, considering the presence of background fields and/or curvature. Even if our flat space analysis shows that gravity is potentially renormalizable at the Planck scale, the presence of fields and/or curvature makes  such a conclusion less evident.
To this purpose, we believe that the role of the spectral dimension itself as an actual spacetime dimension deserves further investigations to include more realistic cases than the mere flat space one.

\begin{acknowledgments}
\noindent P.N. is supported by the Helmholtz International Center for FAIR within the
framework of the LOEWE program (Landesoffensive zur Entwicklung Wissenschaftlich-\"{O}konomischer Exzellenz) launched by the State of Hesse. P.N. would like to thank the Perimeter Institute for Theoretical Physics, Waterloo, ON, Canada for the kind hospitality during the period of work on this project. Research at Perimeter Institute is
supported by the Government of Canada through Industry Canada and by the Province of Ontario through the
Ministry of Research \& Innovation.
\end{acknowledgments}

\end{document}